\documentclass{iaus}
\usepackage{graphicx}

\title[Simulations of Magnetized Winds of Solar-Like Stars]{Numerical Simulations of Magnetized Winds of Solar-Like Stars}

\author[Vidotto et al.]{A. A. Vidotto$^1$$^,$$^2$, M. Opher$^2$, V. Jatenco-Pereira$^1$ \and T. I. Gombosi$^3$}

\affiliation{$^1$University of S\~ao Paulo, Rua do Mat\~ao 1226, S\~ao Paulo, SP, Brazil, 05508-090 \break email: aline@astro.iag.usp.br  \\[\affilskip]
$^2$George Mason University, 4400 University Drive, Fairfax, VA, USA, 22030-4444  \\[\affilskip] 
$^3$University of Michigan, 1517 Space Research Building, Ann Arbor, MI, USA, 48109-2143}

\pubyear{2009}
\volume{259}  
\pagerange{100-100}
\date{Dec 10, 2008 and in revised form ??}
\jname{Cosmic Magnetic Fields: from Planets, to Stars and Galaxies}
\editors{K.G. Strassmeier, A.G. Kosovichev \& J.E. Beckman, eds.}
\begin{document}

\maketitle

\begin{abstract}
We investigate magnetized solar-like stellar winds by means of self-consistent three-dimensional (3D) magnetohydrodynamics (MHD) numerical simulations. We analyze winds with different magnetic field intensities and densities as to explore the dependence on the plasma-$\beta$ parameter. By solving the fully ideal 3D MHD equations, we show that the plasma-$\beta$ parameter is the crucial parameter in the configuration of the steady-state wind. Therefore, there is a group of magnetized flows that would present the same terminal velocity despite of its thermal and magnetic energy densities, as long as the plasma-$\beta$ parameter is the same.
\end{abstract}

\firstsection 

\section{Introduction}
Observations have shown that the solar corona (SC) is a highly complex system, consisting of both long-lived magnetic features (e.g., streamers, coronal holes) and short-lived ones (coronal mass ejections, solar flares, sunspots). Although indirect detections of stellar coronal winds have been performed (Wood et al. 2005), direct measurements of tenuous coronal winds for other stars rather than the Sun are very difficult to do. However, as it occurs in the SC, it is quite certain that the magnetic field is playing an important role in coronal winds of solar-like stars. 

Several studies have been made toward the understanding of an expanding magnetized corona ({\it e.g.,} Pneuman \& Kopp 1971, Low \& Tsinganos 1986, Washimi \& Shibata 1993). However, despite all the notable evolution of both analytical and numerical studies performed in the last decades, we are far from a satisfactory 3D MHD description of a magnetized wind. Several approximations are usually made in order to make the system analytically and numerically tractable.

The present work investigates the influence of the magnetic field in solar-like stellar winds with different $\beta$'s  (the ratio between thermal and magnetic energy densities). We solve the fully 3D MHD equations with the temporal evolution of the energy equation. Therefore, the topology of the field is not restricted and the steady-state arises from the dynamical interplay of the outflow and the field. We neglect the stellar rotation.

\section{The Numerical Model}
To perform the simulations, we make use of BATS-R-US, a 3D ideal MHD numerical code developed at University of Michigan (Powell et al. 1999). We adopted the same, non-uniform grid resolution for all the simulations: the smallest cell size is $0.018~r_0$ (near the central body), the maximum cell size is $4.68~r_0$. The center of the star is placed at the origin of the grid. The axes $x$, $y$, and $z$ extend from $-75~r_0$ to $75~r_0$. 

We consider a star with $1$M$_\odot$, $r_0=1$R$_\odot$. We initialized the simulations with an isothermal HD wind permeating the grid, and a dipolar magnetic field anchored on the stellar surface. The system was then evolved in time until steady-state was achieved. The inner boundary of the system is considered to be the base of the wind at $r_0$, where fixed boundary conditions were adopted. The outer boundary has outflow conditions.

\section{Results}
\begin{figure}
 \includegraphics[scale=0.15]{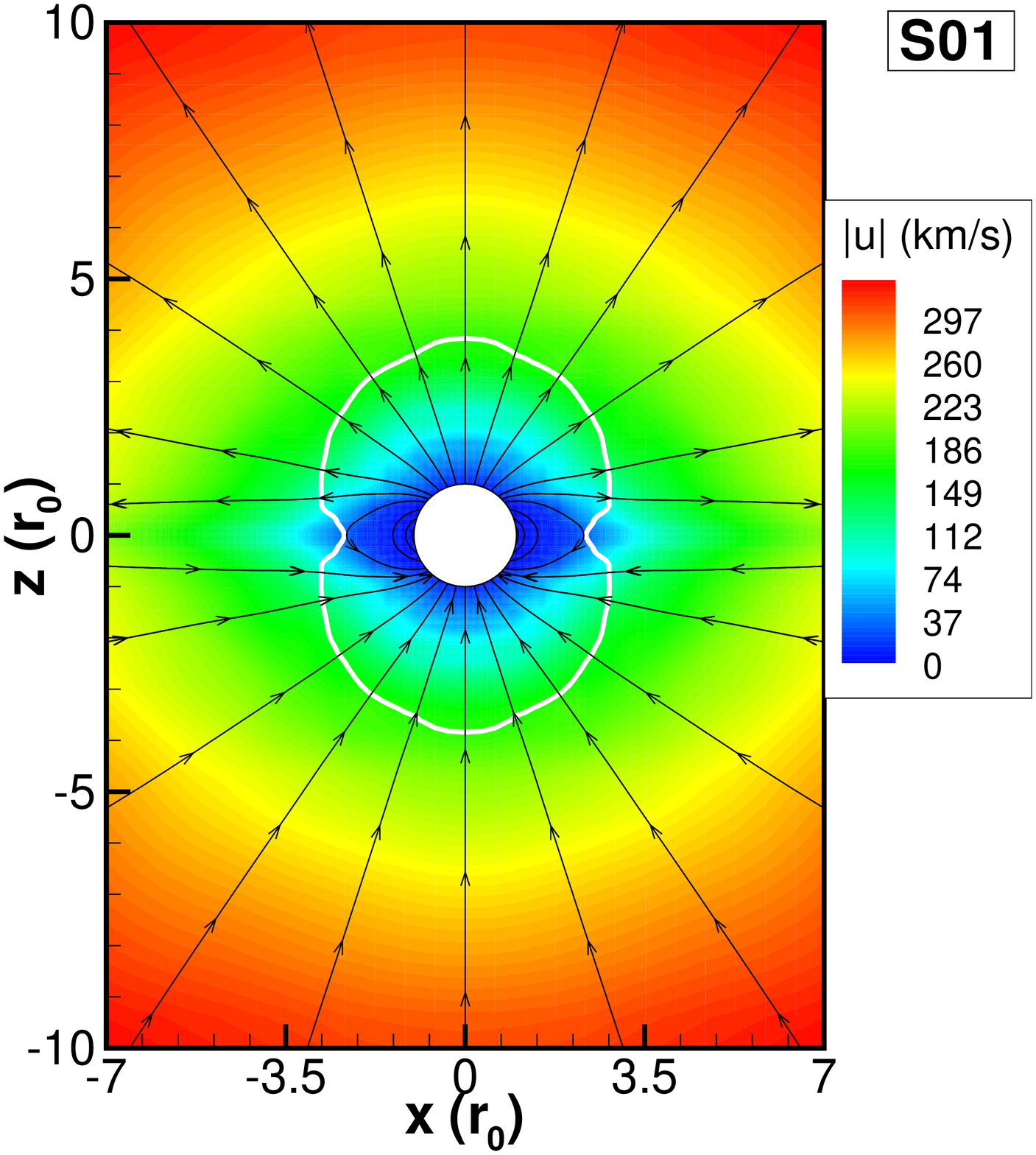}
 \includegraphics[scale=0.15]{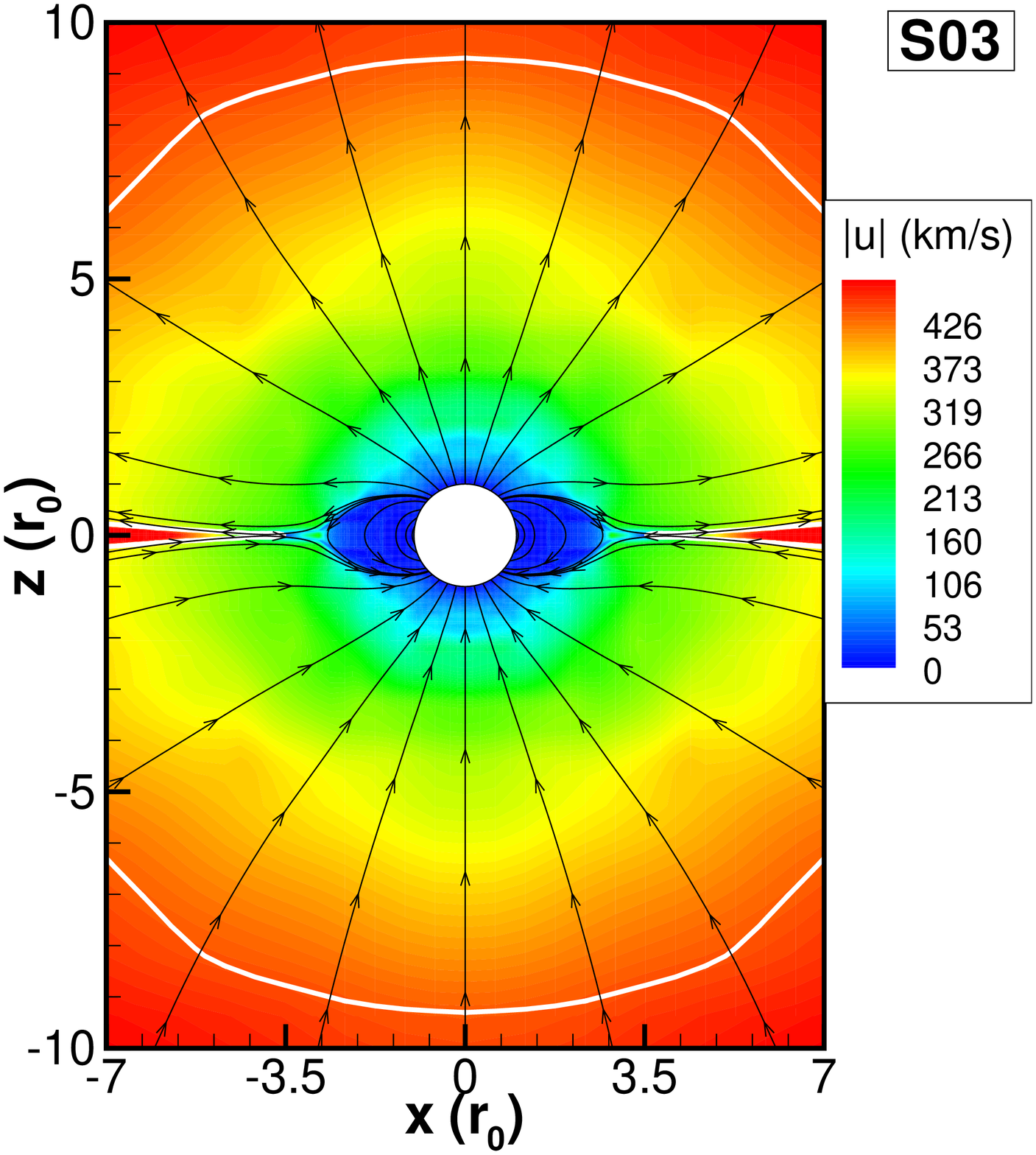}
 \includegraphics[scale=0.15]{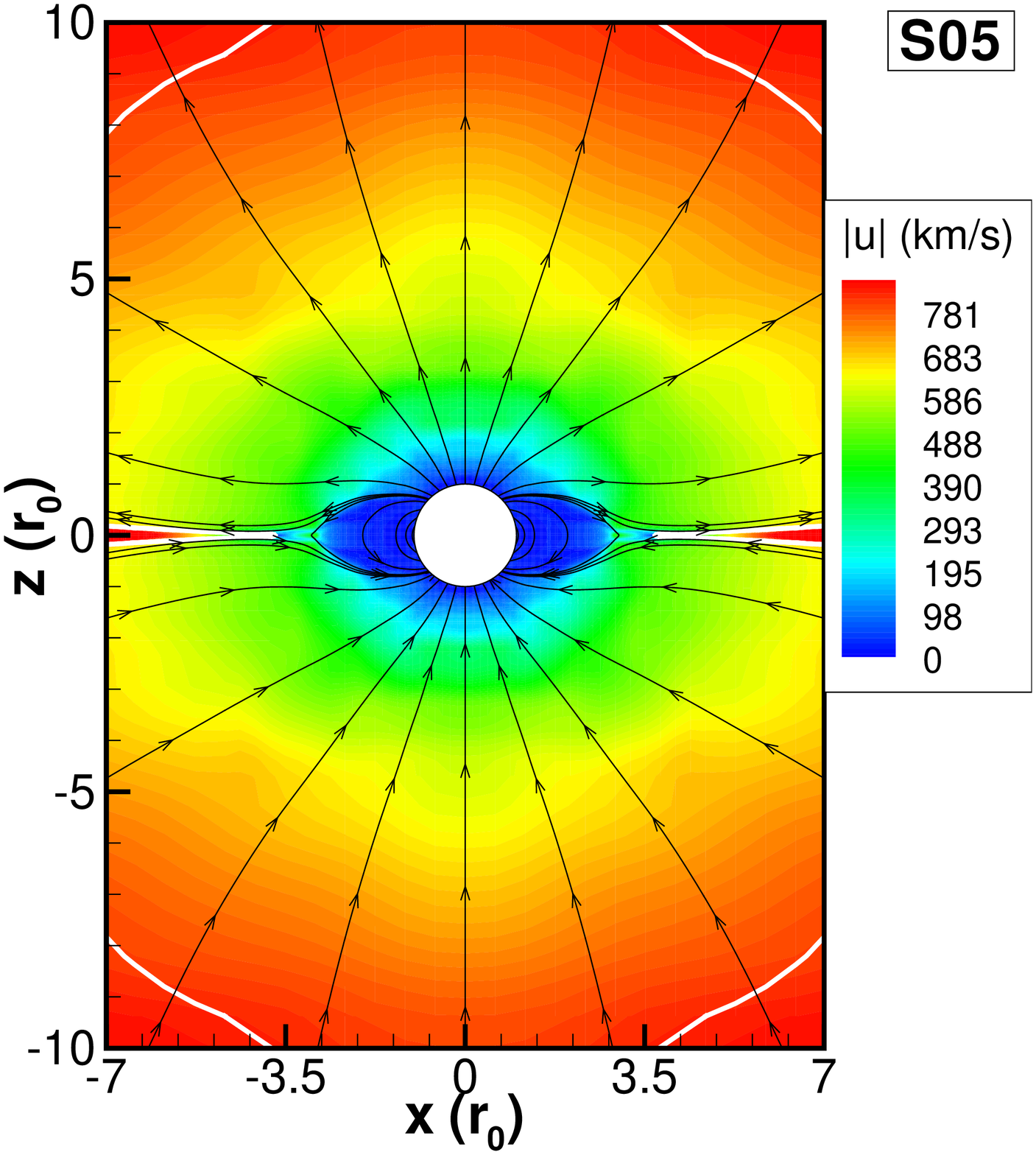}
 \includegraphics[scale=0.15]{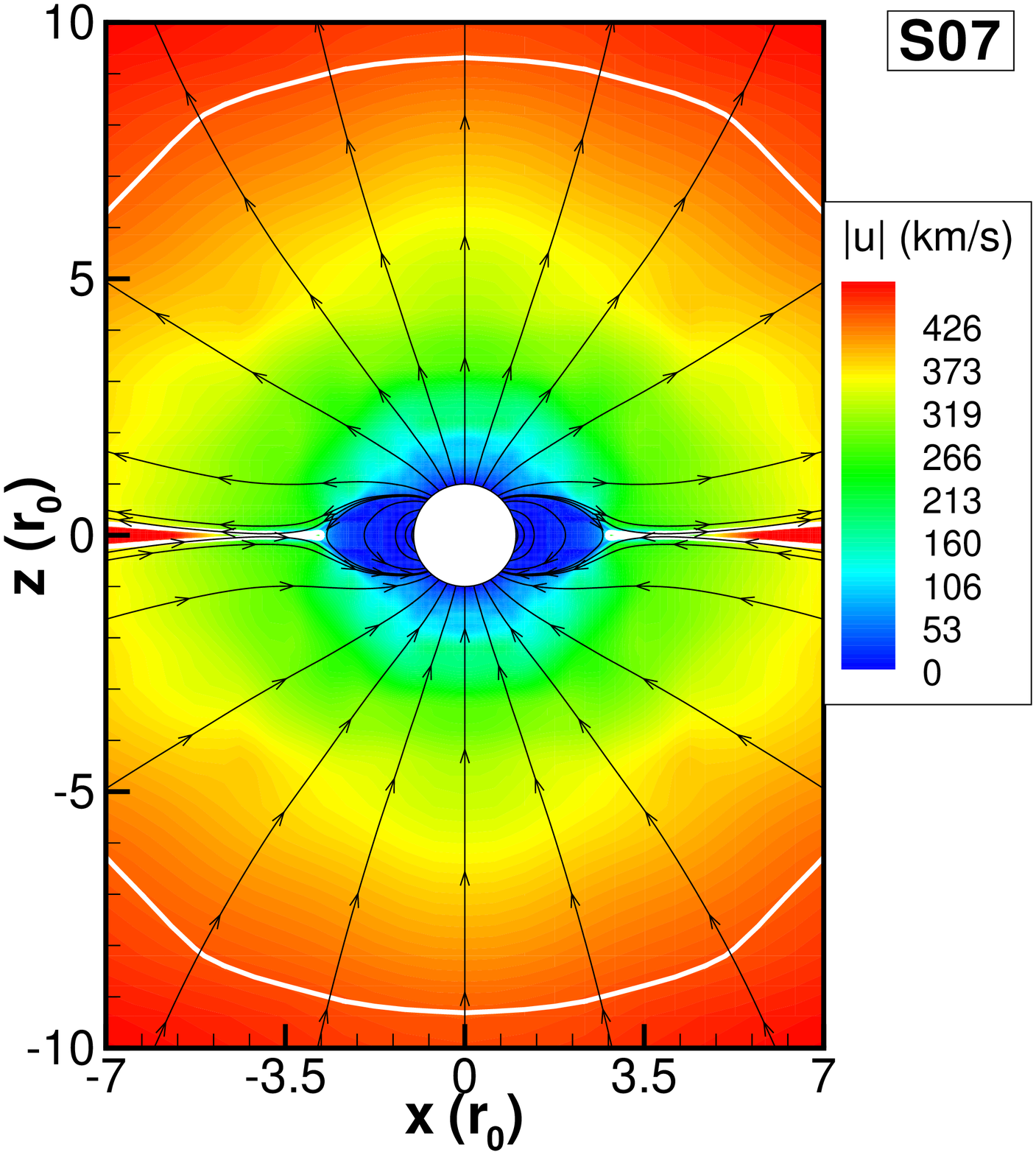}
  \caption{Meridional cuts of the steady-state configurations for cases S01 - S07.}\label{fig:s01-s05}
\end{figure}

We present here four simulations. We consider a corona with temperature and density at the base of the wind of $1.56$~MK and $1.544 \times 10^{-16}$~g~cm$^{-3}$, respectively. The magnitude of the magnetic field was varied in each case. For case S01, $B_0=1$~G; $10$~G in S03; and $20$~G in S05. In case S07, $B_0=20$~G, and density was increased 20 times the previous value. This choice of parameters leads to a plasma-$\beta$ of: $1$ (S01), $0.01$ (S03, S07), and $0.0025$ (S05).  Figure \ref{fig:s01-s05} presents meridional cuts of the steady-state configurations: the contours represent the velocity of the flow, black streamlines are the final configuration of the magnetic field, and the white line is the Alfven surface.

From Figure \ref{fig:s01-s05}, we note that the crucial parameter in determining the wind velocity profile is the $\beta$-parameter (compare S03 and S07). Therefore, there is a group of magnetized flows that would present the same terminal velocity despite of its thermal and magnetic energy densities, as long as the plasma-$\beta$ parameter is the same. We also note that the steady-state magnetic field topology for all cases is similar, presenting a configuration of helmet streamer-type, with zones of closed field lines and open field lines coexisting. The wind is not spherical, it is latitude-dependent: higher velocities are achieved at high latitudes. Cases with lower $\beta$ show more accelerated winds (Fig. \ref{fig:s01-s05}).

The resultant bi-modality of the wind is due to the nature of the magnetic force. A purely HD (non-rotating) wind is spherically symmetric but in the MHD case, this symmetry is lost because the magnetic force has a meridional component.

\bigskip

\noindent{ The authors thank FAPESP (04-13846-6, 07/58793-5), CAPES (BEX4686/06-3), CNPq (305905/2007-4), NSF CAREER (ATM-0747654), and the staff at NASA Ames Research Center for the use of the Columbia supercomputer. }

\end{document}